\documentclass[twocolumn,aps,prl,superscriptaddress]{revtex4-1}
\usepackage{amssymb} 
\usepackage{amsmath,bm} 

\usepackage{graphicx}% Include figure files
\usepackage{dcolumn}% Align table columns on decimal point
\usepackage{bm}% bold math

\usepackage{multirow}
\usepackage[switch,columnwise]{lineno}

\newcommand{\nnu}{\nonumber\\}
\newcommand{\bef}{\begin{figure}[htb]\centering}
\newcommand{\eef}{\end{figure}}

\usepackage[colorlinks=true,linktocpage=true,linkcolor=blue,citecolor=blue]{hyperref}

\graphicspath{{figures/}}

\begin{document}

\title{Twist-4 QCD Factorization for One-Point Correlators at Next-to-Leading Order in Electron-Ion Collisions}% Force line breaks with \\

\author{Omar Elgedawy}
\email[Contact author: ]{omar.elgedawy@polytechnique.edu}
\affiliation{State Key Laboratory of Nuclear Physics and Technology, Institute of Quantum Matter, South China Normal University, Guangzhou 510006, China}
\affiliation{Key Laboratory of Atomic and Subatomic Structure and Quantum Control (MOE), Guangdong-Hong Kong Joint Laboratory of Quantum Matter, Guangzhou 510006, China}
\affiliation{CPHT, CNRS, École polytechnique, Institut Polytechnique de Paris, 91120 Palaiseau, France }

\author{Zhong-Bo Kang}
\email[Contact author: ]{zkang@physics.ucla.edu}
\affiliation{Department of Physics and Astronomy, University of California, Los Angeles, CA 90095, USA}
\affiliation{Mani L. Bhaumik Institute for Theoretical Physics, University of California, Los Angeles, CA 90095, USA}
\affiliation{Center for Frontiers in Nuclear Science, Stony Brook University, Stony Brook, NY 11794, USA}

\author{Hongxi Xing}
\email[Contact author: ]{hxing@m.scnu.edu.cn}
\affiliation{State Key Laboratory of Nuclear Physics and Technology, Institute of Quantum Matter, South China Normal University, Guangzhou 510006, China}
\affiliation{Guangdong Basic Research Center of Excellence for Structure and Fundamental Interactions of Matter, Guangdong Provincial Key Laboratory of Nuclear Science, Guangzhou 510006, China}

\begin{abstract}
Energy and charge correlators have emerged as powerful, infrared- and collinear-safe (IRC-safe) observables for probing QCD dynamics in the vacuum, established within the leading-twist factorization formalism. However, their IRC safety and factorization properties in a nuclear medium, where higher-twist contributions are critically enhanced, remain unproven. In this letter, we present the first proof of next-to-leading order QCD factorization for one-point energy and charge correlators (OPEC and OPCC) in electron-ion collisions at twist-4. Working within the higher-twist expansion framework, we calculate the twist-4 power corrections to the angular widths of the OPEC and OPCC, and explicitly demonstrate that the IRC singularities cancel and fragmentation-function dependence reduces to momentum and charge moments for final-state medium-enhanced observables in semi-inclusive deep inelastic scattering. Supported by numerical evaluations, this framework establishes OPEC and OPCC angular widths as sensitive probes of higher-twist dynamics and fundamental properties of cold nuclear matter.
\end{abstract}

\maketitle

\noindent\textbf{\emph{Introduction}}---Understanding the interactions between energetic partons and nuclear matter is of fundamental importance for high-energy nuclear physics \cite{Wang:2025qdg,Wang:2025lct}. Multiple scatterings of partons traversing a nuclear medium lead to nontrivial phenomena, such as jet quenching and transverse momentum broadening, that encode essential properties of the medium \cite{Gyulassy:1993hr,Baier:1996kr,Gyulassy:2001nm,Zakharov:1997uu,Wang:2001ifa,Guo:2000nz,Vitev:2002pf,Kang:2012kc}. Traditionally, extracting these fundamental medium properties from standard observables like single-hadron transverse momentum distributions or conventional jet shapes has been complicated by their strong dependence on non-perturbative fragmentation processes \cite{Albacete:2013ei,Ru:2019qvz,Ru:2023ars,JETSCAPE:2021ehl,Niida:2021wut,Larkoski:2017jix}. 

To circumvent these non-perturbative limitations, energy and charge correlators have attracted revived interest \cite{Moult:2025nhu}. By tracking the angular distribution of energy or charge flux rather than isolating individual hadrons, these correlators are inherently protected against soft and collinear divergences. In the vacuum, their theoretical framework is highly advanced: rigorous QCD factorization theorems are firmly established, and higher-order perturbative QCD (pQCD) calculations within the leading-twist formalism have enabled an unprecedented precision mapping of vacuum parton evolution in high-energy scattering experiments \cite{CMS:2024mlf,ALICE:2024dfl,STAR:2025jut}. 

Driven by this success, the application of energy correlators has rapidly expanded to probe the properties of quark-gluon plasma and cold nuclear matter in heavy ion collisions~\cite{Andres:2023xwr,Andres:2022ovj,Yang:2023dwc,Devereaux:2023vjz,Liu:2023aqb,Barata:2024wsu,Fu:2024pic}. Experimental collaborations at the LHC and RHIC have recently executed extensive measurements of energy correlators in nuclear collisions \cite{ALICE:2026smt,CMS:2025ydi}, accompanied by a surge of phenomenological modeling and effective theories aiming to describe medium modifications \cite{Fu:2024pic,Fu:2025hpc,Andres:2024xvk,Barata:2024wsu}. However, a critical theoretical gap persists: despite active experimental investigation and phenomenological success, a proof of QCD collinear factorization for energy correlators in the presence of a nuclear medium is missing. Furthermore, the systematic higher-order pQCD calculations required to disentangle intrinsic vacuum radiation from medium-induced scattering remain absent.

Semi-inclusive deep inelastic scattering (SIDIS) offers an exceptionally clean environment to isolate cold nuclear modifications by comparing electron–proton and electron–nucleus collisions, positioning it as a flagship program for the upcoming Electron-Ion Collider (EIC) \cite{AbdulKhalek:2021gbh,Accardi:2012qut}. While the one-point energy and charge correlators (OPEC and OPCC) in SIDIS have been explored primarily within leading-twist collinear or transverse-momentum-dependent (TMD) frameworks \cite{Li:2021txc,Guo:2025qnz,Cao:2023qat,Fu:2025hpc}, probing the structural properties of nuclear matter intrinsically requires systematically incorporating power-suppressed multiple scattering effects \cite{Qiu:1990xxa,Qiu:1990xy}. To fully exploit the EIC program, theoretical predictions must evolve beyond leading-order models, formally integrating these medium effects at higher perturbative orders to validate the underlying factorization theorem.

In this letter, we bridge this fundamental gap by proposing the angular spread of the OPEC and OPCC as a sensitive probe of twist-4 dynamics in cold nuclear matter. We provide the first explicit proof of its QCD factorization at next-to-leading order (NLO) in SIDIS. Utilizing the higher-twist expansion framework, we compute the twist-4 power corrections to the one-point correlators, demonstrating that all medium-associated final-state infrared and collinear divergences safely cancel at NLO. Furthermore, our numerical evaluations demonstrate that the nuclear broadening of these correlators exhibits the characteristic power suppression in $Q^2$ expected of higher-twist QCD dynamics, alongside a linear enhancement with nuclear size ($A^{1/3}$). By expressing the final observable entirely in terms of perturbatively calculable hard coefficients and universal multi-parton quantum correlation functions, this framework elevates the study of medium-modified energy and charge flows from phenomenological modeling to the realm of rigorous QCD factorization.

%========================================================================================
\noindent\textbf{\emph{One-point correlators in SIDIS}}---
To elucidate the role of non-perturbative fragmentation dependencies for one-point correlators, we begin with standard semi-inclusive hadron production off a large nucleus $A$: 
\begin{equation}
e(\ell_1) + A(p) \;\to\; e(\ell_2) + h(p_{h}) + X,
\end{equation}
where $\ell_1$ and $\ell_2$ are the momenta of the incoming and outgoing leptons, respectively. $p$ denotes the average momentum per nucleon in the nucleus of mass number $A$, and $p_{h}$ is the momentum of the observed final-state hadron $h$. In this process, the angular one-point correlators are defined as~\cite{Peccei:1979tc}:
\begin{equation}
    \frac{d\Sigma_{\mathcal{W}}}{d\cos\theta dx_B dy}=\sum_h \int_0^1 dz_h \mathcal{W}_h\frac{d\sigma^h}{d\cos \theta dx_Bdydz_h},
\end{equation}
where the weight $\mathcal{W}_h$ specifies the OPEC ($\mathcal{W}_h=z_h$) or the OPCC
($\mathcal{W}_h=Q_h$), and the sum runs over hadrons produced in the current fragmentation
region~\cite{OPCCDIS}, as described by collinear fragmentation functions. The angle $\theta$ is measured between the out-going hadron and the incoming nucleus in the Breit frame. The standard SIDIS kinematic variables are:
 \begin{equation}
x_B = \frac{Q^2}{2\, p \cdot q}, ~
y = \frac{p \cdot q}{p \cdot \ell_1}, ~
z_h = \frac{p \cdot p_h}{p \cdot q}, ~ 
\hat{x} = \frac{x_B}{x}, ~
 \hat{z} = \frac{z_h}{z},
\end{equation}
where $q=\ell_1-\ell_2$ is the virtual photon momentum and $Q^2=-q^2$. 

To isolate the QCD dynamics of multiple scattering in the nuclear medium, we propose measuring the angular spread of the one-point correlators, defined as:
\begin{equation}
    \langle \sin^2 \theta\rangle_{\Sigma_{\mathcal{W}}} \equiv \frac{\int d\cos \theta \, \sin^2 \theta \frac{d\Sigma_{\mathcal{W}}}{d\cos\theta dx_B dy}}{\int d\cos \theta \, \frac{d\Sigma_{\mathcal{W}}}{d\cos\theta dx_B dy}}.
\end{equation}
Here, we utilize a continuous ${\rm sin}^2\theta$ weighting, which is kinematically related to transverse-momentum-squared weighting for massless partons in the Breit frame. In traditional higher-twist formalisms, such weighting naturally projects out the twist-4 contribution associated with medium-induced transverse momentum broadening. By transitioning to this continuous angular weighting over the inclusive hadronic final-state, we maintain direct sensitivity to medium broadening while constructing a rigorously IRC-safe observable. We emphasize that this approach differs fundamentally from the recently developed light-ray operator product expansion (OPE) framework \cite{Moult:2025nhu}. While light-ray OPE correlators typically track the relative angular substructure between multiple detected particles ($\theta_{ij}$) in the collinear limit, our observable integrates the global angular deflection $\theta$ of the energy/charge flux relative to the incoming collision axis. This distinction seamlessly bridges the standard higher-twist collinear factorization of cold nuclear matter with modern energy and charge correlator measurements.

For single scattering contributions, the leading-twist factorization formalism for this angular spread was established for the OPEC at NLO decades ago~\cite{Peccei:1979kn}. We reproduce this calculation for the OPEC and extend it to the OPCC to demonstrate its IRC safety and factorization properties at leading twist:
\begin{equation}
    \langle \sin^2 \theta\rangle_{\Sigma_{\mathcal{W}}}
    = \frac{\sigma_0\,\frac{\alpha_s}{2\pi}\sum_{q,\bar q} Q_q^2M_q
    \big[f_{q/A}\otimes H_q^{\mathcal{W}}+f_{g/A}\otimes H_g^{\mathcal{W}}\big]}{\Sigma^{(0),T-2}_{\mathcal W}}.
\end{equation}
Here $\Sigma_{\mathcal{W}}^{(0),T-2}$ is the
leading-twist cross section of eq.~(\ref{eq:sigmaW}) and the analytical expressions for the partonic hard coefficients $H_i^{\mathcal{W}}$ are provided in the Supplemental Material.
In the presence of a nuclear medium, multiple scatterings between the energetic parton and nuclear remnants induce nontrivial dynamics, with the leading power corrections arising at twist-4. The QCD factorization for these hadronic scatterings at twist-4 has been extensively developed~\cite{Qiu:1990xxa,Qiu:1990xy} and proven explicitly at NLO for transverse momentum broadening~\cite{Kang:2013raa,Kang:2014ela,Kang:2016ron}. 

For the one-point correlators considered in this letter, we isolate the medium modification through the broadening of the angular spread, defined as the difference between the averaged angular distributions in $e+A$ and $e+p$ collisions: 
\begin{align}
\Delta \langle \sin^2 \theta \rangle_{\Sigma_{\mathcal{W}}} &\equiv 
\langle \sin^2 \theta \rangle_{eA}^{\Sigma_{\mathcal{W}}} - \langle \sin^2 \theta \rangle_{ep}^{\Sigma_{\mathcal{W}}} \nonumber \\ 
&\approx \frac{\int d\cos \theta \, \sin^2 \theta \frac{d\Sigma^{T-4}}{d\cos\theta dx_B dy}}{\int d\cos \theta \, \frac{d\Sigma^{T-2}}{d\cos\theta dx_B dy}} + \dots ,
\label{eq:deltasin2}
\end{align}
where the superscripts $T-4$ and $T-2$ denote the twist-4 and twist-2 contributions, respectively. While higher-twist contributions to the cross section are generally power suppressed by $1/Q^{2}$ in the deep inelastic regime, the twist-4 coefficients are driven by multi-parton nuclear matrix elements that scale with the nuclear size $A^{1/3}$. This geometric enhancement compensates for the kinematic suppression, making twist-4 corrections the dominant medium-induced effect. The omitted terms (dots) represent twist-6 and higher twist contributions, which are safely neglected due to further uncompensated $1/Q^2$ suppression. 

%========================================================================================
\noindent\textbf{\emph{NLO factorization at twist-4}}---Operating within the higher-twist expansion framework developed in Refs.~\cite{Qiu:1990xxa,Qiu:1990xy}, the leading order (LO) contribution at twist-4 can be evaluated straightforwardly  
\begin{align}
&\Delta \langle \sin^2 \theta \rangle^{\rm LO}_{\Sigma_\mathcal{W}} \nonumber \\
=&\frac{16\pi^2\alpha_s\sigma_0}{N_c Q^2\Sigma_{\mathcal{W}}^{(0),T-2}}\sum_{h,q} Q_q^2 \int_0^1 dz_h\, \mathcal{W}_h\, T_{qg}(x_B,0,0)\,D_{q}^h(z_h) \nonumber\\
=&\frac{16\pi^2\alpha_s}{N_c Q^2}\,
\frac{\sum_q Q_q^2 M_q\, T_{qg}(x_B,0,0)}{\sum_q Q_q^2 M_q\, f_{q/A}(x_B)},
\label{eq:t4-lo}
\end{align}
where $\sigma_0=2\pi\alpha_{\rm em}^2[1+(1-y)^2]/(yQ^2)$, $\Sigma_{\mathcal{W}}^{(0),T-2}$ is the leading twist cross section evaluated at LO in the second line,
and the sums run over quarks and antiquarks. By virtue of energy conservation $M_q=1$ for the
OPEC, while charge conservation gives $M_q=Q_q$ and $M_{\bar q}=-Q_q$ for the OPCC. Possible nonperturbative modifications of these moments~\cite{Collins:2023cuo,Ke:2026zia} affect the charge weighting but do not alter the perturbative factorization structure established here.

The LO expression depends on fragmentation functions only through their momentum or charge moments, given by the following sum rules:
\begin{equation}
    \sum_h \int_0^1 dz_h \mathcal{W}_h D_{i}^h(z_h)=M_i.
    \label{eq-sumrule}
\end{equation}
This highlights the unique capability of one-point correlators to probe the fundamental structure of the nuclear medium. The dynamics are entirely encapsulated within the twist-4 matrix element, $T_{qg}$, which represents the quantum multi-parton correlation of quarks and gluons inside the nucleus, defined as: \cite{Luo:1992fz,Luo:1994np}
\begin{align}
T_{qg}(x_1, x_2, x_3)
&= \int \frac{dy^-}{2\pi} e^{ix_1p^+y^-}  \int \frac{dy_1^-dy_2^-}{4\pi} 
\nnu
&\hspace{-24pt}\times e^{ix_2p^+(y_1^- - y_2^-)}
 e^{ix_3p^+y_2^-} \theta(y_2^-)\theta(y_1^- - y^-)
\nnu
&\hspace{-24pt}\times  \langle A|{\bar\psi}_q(0) \gamma^+ F_{\sigma}^+(y_2^-)F^{\sigma +}(y_1^-)\psi_q(y^-)|A\rangle.
\end{align}

\begin{figure}
    \centering
    \includegraphics[width=0.9\linewidth]{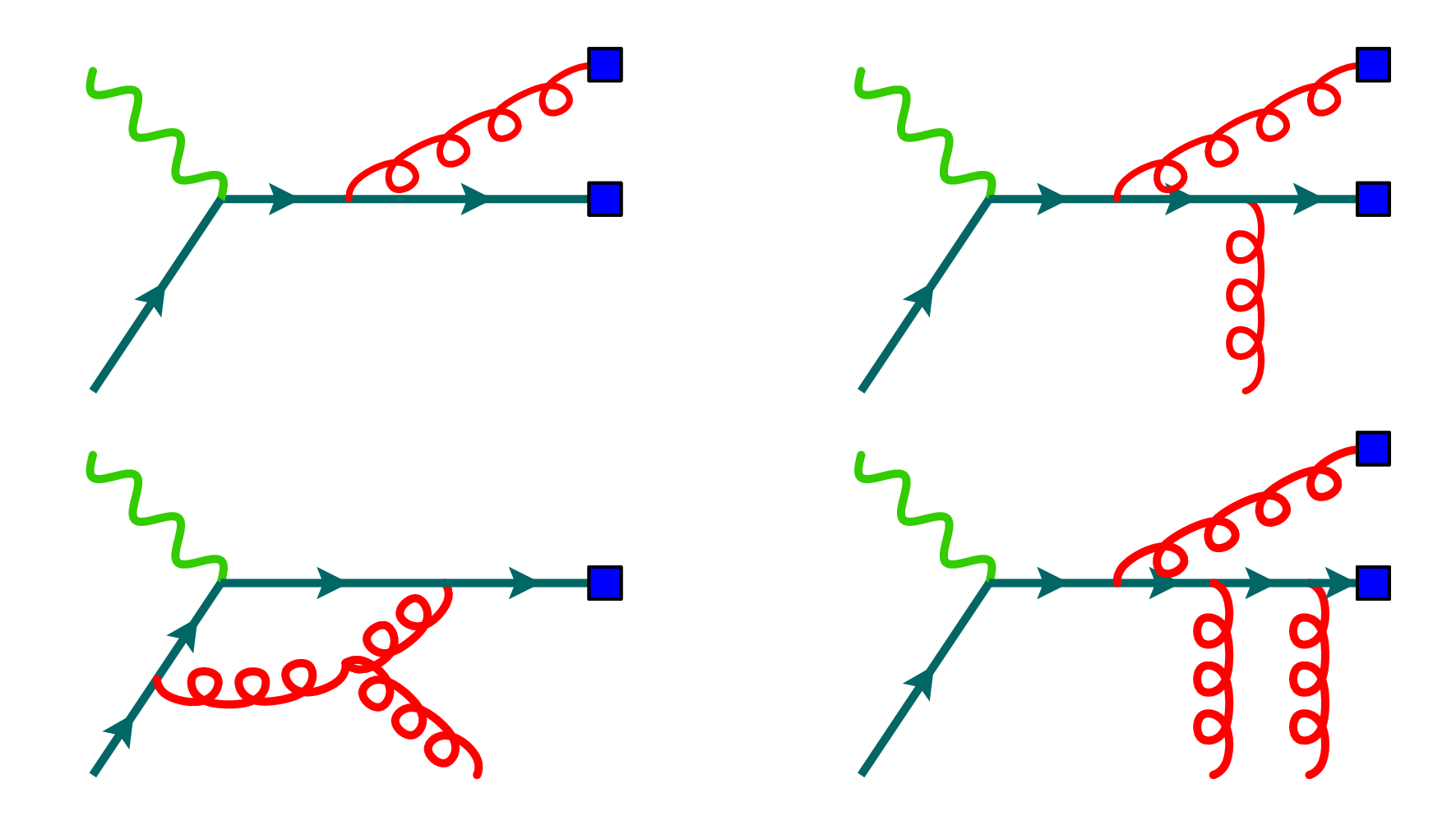}
    \caption{Sample Feynman diagrams at NLO from single scattering (top-left), double scattering real correction (top-right), double scattering virtual correction (bottom-left) and triple scattering (bottom-right). The blue squares represent energy or charge measurements.}
    \label{fig:feyndiagm}
\end{figure}

Establishing the IRC safety of one-point correlators at NLO is critical. At twist-4, this requires evaluating double-scattering contributions arising from both hard and soft rescatterings, along with their interferences. Some representative diagrams are shown in fig.~\ref{fig:feyndiagm}. For the NLO virtual corrections at twist-4, evaluating the diagrams in dimensional regularization ($d=4-2\epsilon$) yields:  
\begin{align}
\Delta \langle \sin^2 \theta \rangle_{\Sigma_\mathcal{W}}^{\rm NLO\text{-}V}
&= \Delta \langle \sin^2 \theta \rangle_{\Sigma_\mathcal{W}}^{\rm LO}\,
\frac{\alpha_s}{2\pi}\left(\frac{4\pi\mu^2}{Q^2}\right)^{\epsilon}\nonumber\\
&\times\frac{C_F}{\Gamma(1-\epsilon)}
\left(-\frac{2}{\epsilon^2}-\frac{3}{\epsilon}-8\right). 
\label{eq:t4-nlo-v}
\end{align}
As with the LO case, the FFs decouple completely upon applying the corresponding sum rules.

The evaluation of twist-4 real corrections at NLO is more involved, encompassing double-scattering contributions and interferences between single and triple scatterings. We find that the single-triple scattering interference is completely free of infrared and collinear singularities. The double-scattering contributions, however, contain both:
\begin{align}
&\Delta \langle \sin^2 \theta \rangle_{\Sigma_\mathcal{W}}^{\rm NLO\text{-}R} 
={}\frac{16\pi^2\alpha_s\,\sigma_0}{N_c Q^2\,\Sigma_{\mathcal{W}}^{(0)}}\,\frac{\alpha_s}{2\pi}
\left(\frac{4\pi\mu^2}{Q^2}\right)^{\epsilon}\frac{1}{\Gamma(1-\epsilon)}
\nonumber\\
&\times\sum_{q}Q_q^2\sum_{h,j}\int_0^1 dz_h\, \mathcal{W}_h
\int_{x_B}^1\frac{dx}{x}\int_{z_h}^1\frac{dz}{z}\,D_{j}^h(z)
\nonumber\\
&\times\Big[\frac{2C_F}{\epsilon^2}\,T_{qg}(x,0,0)\,\delta_{\hat x}\delta_{\hat z}
-\frac{1}{\epsilon}\,\delta_{\hat z}\, I_{j}(x)-\frac{1}{\epsilon}\,\delta_{\hat x}\,F_{j}(x,\hat z)
\nonumber\\
&\quad+H_{j}\otimes T_{qg}\Big],
\label{eq:t4-nlo}
\end{align}
where $\delta_{\hat x}=\delta(1-\hat x)$ and $\delta_{\hat z}=\delta(1-\hat z)$. Notice that the soft-collinear singularity (the $1/\epsilon^2$ term) perfectly cancels in the sum of the real and virtual corrections. $H_j$ denotes the finite part. The term proportional to $I_j$ isolates the initial-state divergence; because it is independent of the momentum fraction $z$, one can directly apply the sum rules shown in eq.~(\ref{eq-sumrule}) to eliminate the FFs. The remaining initial-state divergence is then absorbed into the renormalization of the twist-4 matrix element $T_{qg}$, yielding its DGLAP-like QCD evolution~\cite{Kang:2013raa,Kang:2014ela,Kang:2016ron}. The term $F_j$ represents the final-state divergence, capturing both quark and gluon fragmentation channels:
\begin{align}
F_{q} =& F_{q}^{hh} +F_{q}^{ss}+F_{q}^{sh}+F_{q}^{hs}+F_{q}^{\rm virt} 
= T_{qg}(x,0,0) P_{qq}(\hat z_q),       \\
F_{g} =& F_{g}^{hh} +F_{g}^{ss}+F_{g}^{sh}+F_{g}^{hs} 
= T_{qg}(x,0,0) P_{gq}(\hat z_g).
\end{align}
Here, $F_{q/g}^{ab}$ distinguish the soft and hard rescattering components (detailed in the Supplemental Material). $P_{ij}(z)$ are the standard DGLAP splitting kernels for partons $j\rightarrow i$:
\begin{align}
P_{qq}(z) &= C_F \left[ \frac{1+z^2}{(1-z)_{+}} + \frac{3}{2}\,\delta(1-z) \right], \label{eq:Pqq}\\[6pt]
P_{gq}(z) &= C_F \,\frac{1+(1-z)^2}{z}.
\label{eq:Pgq}
\end{align}
Summing the contributions from quark and gluon energy detection, the total final-state divergence of the one-point correlators at twist-4 NLO is:
\begin{align}
  & \sum_{h,j}\int_0^1 dz_h\int_{z_h}^1\frac{dz}{z}\,\mathcal{W}_h\, D_j^h(z)\,F_j(x_B,\hat z) \nonumber \\
  &=T_{qg}(x_B,0,0)\int_0^1 d\hat{z}\;\hat{\mathcal{W}}(\hat z)
  \Big[M_q\,P_{qq}(\hat{z})+M_g\,P_{gq}(\hat{z})\Big]\nonumber \\
  &=0,
    \label{eq-divT4}
\end{align}
where we changed variables to $\hat z=z_h/z$ and used the sum rules, with
$\hat{\mathcal W}(\hat z)=\hat z$ and $M_q=M_g=1$ for the OPEC, and
$\hat{\mathcal W}(\hat z)=1$, $M_q=Q_q$, $M_g=0$ for the OPCC. We also used
\begin{align}
    \int_0^1 dz\,z \left[P_{qq}(z)+ P_{gq}(z)\right]=0,
    \qquad
    \int_0^1 dz\,P_{qq}(z)=0.
\end{align}
This result explicitly demonstrates the exact cancellation of all final-state soft and collinear singularities, rigorously proving the IRC safety of one-point correlators at twist-4. While the correlators carried individually by quarks or gluons are not separately infrared-safe, their inclusive sum is perfectly finite. 

Furthermore, we can explicitly show that the fragmentation-function dependence of the remaining finite part reduces to the same moments:
\begin{align}
  & \sum_{h}\int_0^1 dz_h\int^1_{z_h}\frac{dz}{z}\mathcal{W}_h D_j^h(z)H_j(\hat x,\hat z) 
  =M_j\tilde{H}_j(\hat x),
    \label{eq:Hj}
\end{align}
where the analytical expressions for the finite parts $\tilde{H}_j$ for each channel are provided in Eqs. (\ref{eq:Htelda}) and (\ref{eq:Htelda_opcc}) of the Supplemental Material. This leads to the final, purely perturbative NLO QCD factorization formula for one-point correlators:
\begin{align}
\Delta \langle \sin^2 \theta \rangle_{\Sigma_{\mathcal{W}}}^{\rm NLO} 
=&\frac{16\pi^2\alpha_s\,\sigma_0}{N_c Q^2\,\Sigma_{\mathcal{W}}^{(0)}}\,\frac{\alpha_s}{2\pi}
\sum_{q}Q_q^2\sum_{j}M_j
\nonumber\\
&\times\int_{x_B}^1\frac{dx}{x}
\left[\tilde{H}_{j}(\hat x)\otimes T_{qg}  \right],  
\label{eq:t4-nlo-fact}
\end{align}
where $\Sigma_{\mathcal{W}}^{(0)}$ is evaluated at NLO, and the sum over the contributions with the different twist-4 matrix elements is implicit and given in the supplemental material.

%=====================================================
\noindent\textbf{\emph{Numerical Results}}--- To quantitatively estimate the twist-4 corrections, we focus on the soft rescattering contributions. The corresponding diagonal twist-4 matrix elements are relatively well-constrained by existing experimental data, whereas the off-diagonal matrix elements required for hard rescatterings remain poorly known and are deferred to future investigations. Assuming a large, loosely bound nucleus where the spatial and momentum correlations
between two nucleons can be safely neglected, the diagonal matrix elements can be
effectively factorized as
\begin{equation}
T_{ig}(x,0,0;\mu^2)
\;\approx\;
\frac{9\,R_A}{8\pi^2\,\alpha_s}\,
f_{i/A}(x,\mu^2)\,
\hat{q}_0,
\qquad i=q,g,
\label{eq:Tqg_factorized}
\end{equation}
where $R_A\approx 1.12\,A^{1/3}$~fm is the nuclear radius and $f_{i/A}$ are the nuclear
PDFs, for which we use the EPPS21 parametrization~\cite{nPDF}. The parameter $\hat{q}_0$ represents the nuclear-geometry-averaged quark jet transport coefficient, which we assume to be independent of $A$. Consistent with global extractions in cold nuclear matter, we adopt $\hat{q}_{\rm 0} \approx 0.02~\mathrm{GeV^2/fm}$~\cite{Ru:2019qvz}, which is approximately two orders of magnitude smaller than its typical values in a hot quark-gluon plasma. We neglect any potential scale dependence of \(\hat{q}\), treating it as a constant over the considered range of Bjorken \(x_B\) and momentum transfer \(Q^2\). All predictions are computed for $e+A$ collisions at $\sqrt s=89$~GeV.

\begin{figure}
    \centering
    \includegraphics[width=1\linewidth]{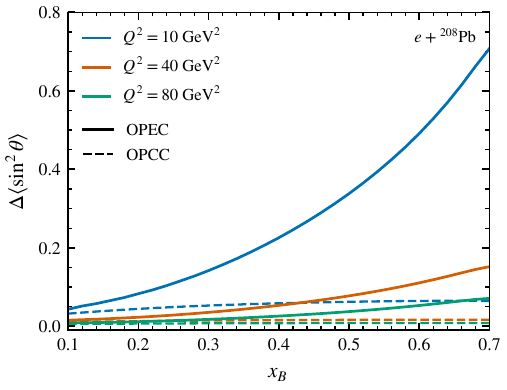}
   \caption{Twist-4 nuclear angular spread $\Delta\langle\sin^2\theta\rangle$ of the
one-point energy correlator (OPEC, solid) and charge correlator (OPCC, dashed) as a
function of $x_B$ in $e+{}^{208}$Pb collisions at $\sqrt{s}=89$~GeV, for
$Q^2=10$, $40$ and $80$~GeV$^2$. We use EPPS21 nuclear PDFs,
$\mu=Q$, $\hat q_0=0.02$~GeV$^2$/fm and $R_A=1.12\,A^{1/3}$~fm.}
\label{fig:xB}
\end{figure}

In Fig.~\ref{fig:xB}, we present the angular width $\Delta\langle \rm \sin^2\theta\rangle$ of the OPEC (solid lines) and OPCC (dashed lines) at twist-4 as a function of $x_B$ for various fixed values of $Q^2$ in a lead nucleus. The observables exhibit a strong, multidimensional dependence on both $x_B$ and the probing scale $Q^2$. Crucially, the twist-4 broadening is systematically suppressed as $Q^2$ increases, perfectly mirroring the power-suppressed behavior dictated by the QCD twist expansion. Furthermore, it is evident that the overall magnitude of the OPCC broadening is systematically smaller than that of the OPEC. This distinct suppression intrinsically arises from the electric charge weighting defining the OPCC; most significantly, gluons carry zero electric charge, which completely decouples the gluon-fragmentation and gluon-initiated channels that otherwise strongly enhance the energy-weighted OPEC.

\begin{figure}
    \centering
    \includegraphics[width=1\linewidth]{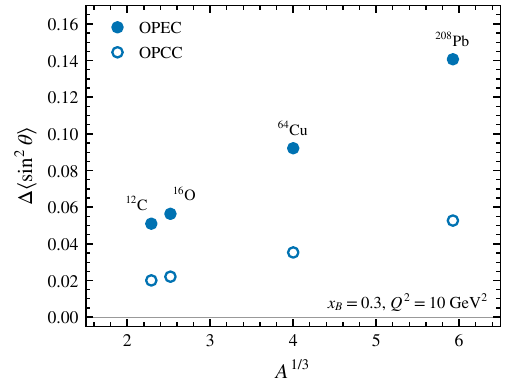}
    \caption{Nuclear-size dependence of the twist-4 angular spread
$\Delta\langle\sin^2\theta\rangle$ for the OPEC (filled circles) and OPCC (open circles)
at $x_B=0.3$ and $Q^2=10$~GeV$^2$, for $e+{}^{12}$C, ${}^{16}$O, ${}^{64}$Cu and
${}^{208}$Pb collisions at $\sqrt{s}=89$~GeV.}
\label{fig:A13fig}
\end{figure}
In Fig.~\ref{fig:A13fig}, we display the angular width of the OPEC and OPCC as a function of $A^{1/3}$ for $^{12}\rm C$, $^{16}\rm O$, $^{64}\rm Cu$, and $^{208}\rm Pb$ at fixed kinematics ($x_B = 0.3$ and $Q^2 = 10$ GeV$^2$). The broadening grows linearly with $A^{1/3}$, directly reflecting the geometric enhancement embedded in the twist-4 quark-gluon correlation function ($T_{qg} \propto R_A \hat q_0$). This robust linear scaling confirms that the angular spread of one-point correlators provides a highly sensitive, definitive signature of higher-twist nuclear dynamics.

%====================================================================
\noindent\textbf{\emph{Summary}}---In this Letter, we presented the first explicit proof of next-to-leading order (NLO) QCD factorization for the twist-4 contributions to the angular widths of the OPEC and OPCC in SIDIS. We demonstrated that these observables are strictly infrared- and collinear-safe; all medium-associated final-state singularities identically cancel, and fragmentation-function dependence reduces to momentum and charge moments. Supported by our numerical evaluations of soft rescattering dynamics, we established that the medium-induced broadening of these correlators exhibits the characteristic $Q^2$ suppression of higher-twist QCD, alongside a robust linear scaling with nuclear size ($A^{1/3}$). 

By establishing these angular spreads as probes of nuclear multiple scattering, this framework provides a rigorous theoretical baseline for extracting fundamental multi-parton correlations in cold nuclear matter at the upcoming Electron-Ion Collider (EIC). Furthermore, successfully formally integrating higher-twist medium effects into a collinear factorization theorem represents a critical theoretical stepping stone. It lays the groundwork for systematically extending higher-order perturbative QCD calculations of energy correlators to heavy-ion collisions, ultimately promising unprecedented precision in the tomographic study of the quark-gluon plasma.

\begin{acknowledgments}
The work of H.X. was supported by National Natural Science Foundation of China with Projects No. 12525508 and No. 12475139. The work of O.E. was supported at different times by the National Natural Science Foundation of China under project No. 12447145 and by the European Union (ERC, QGPthroughEECs, grant agreement No. 101164102). Views and opinions expressed are however those of the authors only and do not necessarily reflect those of the European Union or the European Research Council. Neither the European Union nor the granting authority can be held responsible for them. 
Z.K. is supported by the National Science Foundation under grant No.~PHY-2515057.
\end{acknowledgments}

\bibliographystyle{h-physrev5}   
\bibliography{refs.bib}
%===============================
% Supplemental Material section
%===============================
\clearpage
\onecolumngrid
\setcounter{secnumdepth}{2}          % number (sub)sections in the SM only
\setcounter{section}{0}
\renewcommand{\thesubsection}{S\arabic{subsection}}
\begin{center}
\textbf{\large Supplemental Material}
\end{center}
\setcounter{equation}{0}
\setcounter{figure}{0}
\setcounter{table}{0}
\renewcommand{\theequation}{S\arabic{equation}}
\renewcommand{\thefigure}{S\arabic{figure}}
\renewcommand{\thetable}{S\arabic{table}}

\subsection{Angular-weighted OPCC at leading twist (NLO)}
\label{sec:angular_opcc}

At leading twist, the angular spread defined in the main text is the ratio
\begin{equation}
\left\langle\sin^2\theta\right\rangle^{\Sigma_{\mathcal{W}}}
=\frac{\Sigma_{\mathcal{W}}^{(w),T-2}}{\Sigma_{\mathcal{W}}^{(0),T-2}},
\label{eq:angular_spread}
\end{equation}
where
\begin{align}
\Sigma_{\mathcal{W}}^{(w)}&\equiv\int d\cos\theta\,\sin^2\theta\,
\frac{d\Sigma_{\mathcal{W}}}{d\cos\theta\,dx_B\,dy},&
\Sigma_{\mathcal{W}}^{(0)}&\equiv\int d\cos\theta\,
\frac{d\Sigma_{\mathcal{W}}}{d\cos\theta\,dx_B\,dy},
\label{eq:sigmaW}
\end{align}
and the superscript $T-2$ denotes the twist-2 contribution. Both
quantities are differential in $x_B$ and $y$ only. With $\theta$ the
Breit-frame angle defined in the main text, the angular weight in terms
of the partonic variables $\hat x=x_B/x$ and $\hat z=z_h/z$ is
\begin{equation}
w(\hat x,\hat z)\equiv\sin^2\theta
=\frac{4\hat x\hat z(1-\hat x)(1-\hat z)}{(1-\hat x-\hat z+2\hat x\hat z)^2}.
\label{eq:angular_weight}
\end{equation}
We also use $\kappa_y\equiv\frac{1+4(1-y)+(1-y)^2}{1+(1-y)^2}$ and
$L\equiv\ln\frac{1-\hat x}{\hat x}$.

We denote by $H^{T-2}_{ij}(\hat x,\hat z)$ the NLO twist-2 partonic hard
coefficient for an incoming parton $i$ and a fragmenting parton $j$, taken
from Ref.~\cite{Kang:2014ela}. The splitting kernels $P_{ij}$ ($j\to i$) are
those of Eqs.~\eqref{eq:Pqq} and \eqref{eq:Pgq}, supplemented by
$P_{qg}(z)=T_R\left[z^2+(1-z)^2\right]$. The leading-twist hard
coefficients $H_i^{\mathcal{W}}$ of the main text come in two versions,
$H_i^{(0),\mathcal{W}}(\hat x)$ for the unweighted and
$H_i^{(w),\mathcal{W}}(\hat x)$ for the angular-weighted cross section.

We first consider the one-point charge correlator (OPCC),
$\mathcal{W}_h=Q_h$. Applying the sum rule of eq.~\eqref{eq-sumrule} gives
$M_q=Q_q$, $M_{\bar q}=-Q_q$ and $M_g=0$. This removes
gluon-fragmentation and gluon-initiated channels, so only the channel
$f_{q/A}D_q^h$ survives, and the sums below run over quarks and
antiquarks.

For the unweighted OPCC we define
\begin{equation}
H^{(0),Q_h}_{q}(\hat x)\equiv\int_0^1 d\hat z\,
H^{T-2}_{qq}(\hat x,\hat z),
\end{equation}
which evaluates to
\begin{align}
H^{(0),Q_h}_{q}(\hat x)
={}&C_F\Bigg\{\Big(-\frac92-\frac{\pi^2}{3}\Big)\delta(1-\hat x)
+\frac{\frac12-2\hat x}{(1-\hat x)_+}
+(1+\hat x^2)\Big(\frac{\ln(1-\hat x)}{1-\hat x}\Big)_+
-\frac{1+\hat x^2}{1-\hat x}\ln\hat x\nonumber\\
&\qquad+(1-\hat x)+\kappa_y\,\hat x\Bigg\}.
\label{eq:hbarfinal}
\end{align}
The final-state logarithm vanishes after the $\hat z$ integration,
since $\int_0^1 d\hat z\,P_{qq}(\hat z)=0$, so
\begin{align}
\Sigma_{Q_h}^{(0),T-2}
={}&\sigma_0\sum_{q,\bar q}Q_q^2M_q\int_{x_B}^1\frac{dx}{x}\,f_{q/A}(x,\mu^2)
\Bigg[\delta(1-\hat x)+\frac{\alpha_s}{2\pi}\Big(
\ln\frac{Q^2}{\mu^2}P_{qq}(\hat x)+H^{(0),Q_h}_{q}(\hat x)\Big)\Bigg].
\label{eq:opcc_unweighted_explicit}
\end{align}

For the angular-weighted OPCC the relevant coefficient is
\begin{align}
H^{(w),Q_h}_{q}(\hat x)&\equiv\int_0^1 d\hat z\,w(\hat x,\hat z)\,
H^{T-2}_{qq}(\hat x,\hat z)\nonumber\\
&=C_F\Bigg\{-\frac{2}{(2\hat x-1)^4}\Bigg[2\hat x(4\hat x^4-4\hat x^3+1)\,L
+16\hat x^5-56\hat x^4+68\hat x^3-36\hat x^2+11\hat x-2\Bigg]\nonumber\\
&\qquad+\kappa_y\,\frac{4\hat x^2(\hat x-1)}{(2\hat x-1)^4}
\Bigg[-2(1-\hat x^2)\,L+8\hat x^2-14\hat x+5\Bigg]\Bigg\}.
\label{eq:hbar_weighted_def}
\end{align}
The apparent poles at $\hat x=1/2$ cancel between the logarithmic and
rational terms. Since $w(\hat x,1)=w(1,\hat z)=0$, the Born term and both
explicit collinear logarithms vanish, and
\begin{equation}
\Sigma_{Q_h}^{(w),T-2}=\sigma_0\frac{\alpha_s}{2\pi}\sum_{q,\bar q}Q_q^2M_q
\int_{x_B}^1\frac{dx}{x}\,f_{q/A}(x,\mu^2)\,H^{(w),Q_h}_{q}(\hat x).
\label{eq:opcc_weighted}
\end{equation}
The angular spread of the OPCC is then
\begin{equation}
\left\langle\sin^2\theta\right\rangle^{\Sigma_{Q_h}}
=\frac{\Sigma_{Q_h}^{(w),T-2}}{\Sigma_{Q_h}^{(0),T-2}}.
\label{eq:angular_spread_opcc}
\end{equation}

\subsection{Angular-weighted OPEC at leading twist (NLO)}
\label{sec:angular_opec}

We now repeat the calculation for 
one-point energy correlator (OPEC), $\mathcal{W}_h=z_h$, with the same
weight $w(\hat x,\hat z)$.

\paragraph{Sum rules and partonic structure.}
Starting from eq.~\eqref{eq:sigmaW}, and inserting the weight
$z_h=\hat z\,z$ with the measure $dx\,dz/(xz)$, the hadron-level moment
becomes
\begin{equation}
\int dz_h\,z_h\int\frac{dz}{z}\,D_{i}^h(z)\,F(z_h/z)
=\int dz\,z\,D_{i}^h(z)\int_0^1 d\hat z\,\hat z\,F(\hat z).
\end{equation}
Applying eq.~\eqref{eq-sumrule} with $\mathcal{W}_h=z_h$ gives $M_i=1$ for
$i=q,\bar q,g$, so one explicit factor $\hat z$ remains in the partonic
convolution. Since the sum rule holds for every parton, the
gluon-initiated and gluon-fragmentation channels survive. We define
\begin{align}
H^{(0),z_h}_{q}(\hat x)&\equiv\int_0^1 d\hat z\,\hat z\,
\Big[H^{T-2}_{qq}+H^{T-2}_{qg}\Big],&
H^{(0),z_h}_{g}(\hat x)&\equiv\int_0^1 d\hat z\,\hat z\,
H^{T-2}_{gq},\\
H^{(w),z_h}_{q}(\hat x)&\equiv\int_0^1 d\hat z\,\hat z\,w
\Big[H^{T-2}_{qq}+H^{T-2}_{qg}\Big],&
H^{(w),z_h}_{g}(\hat x)&\equiv\int_0^1 d\hat z\,\hat z\,w\,
H^{T-2}_{gq}.
\end{align}
In the gluon-initiated channel, the sum over $q,\bar q$ below
automatically includes both the quark and the antiquark fragmentation
channels.

\paragraph{Unweighted OPEC.}
The final-state collinear logarithm cancels after summing over the
fragmenting parton, since
$\int_0^1 d\hat z\,\hat z\,P_{qq}(\hat z)=-\tfrac43C_F$ and
$\int_0^1 d\hat z\,\hat z\,P_{gq}(\hat z)=+\tfrac43C_F$.
The two quark-initiated contributions are
\begin{align}
\int_0^1 d\hat z\,\hat z\,H^{T-2}_{qq}
={}&C_F\Bigg\{\Big(-\frac{19}{9}-\frac{\pi^2}{3}\Big)\delta(1-\hat x)
-\frac{6\hat x^2+6\hat x+5}{6\,(1-\hat x)_+}
+(1+\hat x^2)\Big(\frac{\ln(1-\hat x)}{1-\hat x}\Big)_+\nonumber\\
&\quad-\frac{1+\hat x^2}{1-\hat x}\ln\hat x+(1-\hat x)
+\frac23\,\kappa_y\,\hat x\Bigg\},\\
\int_0^1 d\hat z\,\hat z\,H^{T-2}_{qg}
={}&C_F\Bigg\{-\frac{43}{18}\,\delta(1-\hat x)
+\frac{3\hat x^2-3\hat x+4}{3\,(1-\hat x)_+}
+\frac13\,\kappa_y\,\hat x\Bigg\},
\end{align}
and their sum is
\begin{equation}
H^{(0),z_h}_{q}(\hat x)=H^{(0),Q_h}_{q}(\hat x).
\end{equation}
This equality is not a property of the integrands:
$\int d\hat z\,H^{T-2}_{qq}$ counts quark number, while
$\int d\hat z\,\hat z\,[H^{T-2}_{qq}+H^{T-2}_{qg}]$ counts momentum, and
both reduce to the inclusive quark coefficient function. The
gluon-initiated coefficient is
\begin{equation}
H^{(0),z_h}_{g}(\hat x)=\big(L-1\big)\,P_{qg}(\hat x)
+T_R\,\kappa_y\,2\hat x(1-\hat x).
\end{equation}
For $\kappa_y=3$ ($y\to0$), $H^{(0),z_h}_{q}$ and $H^{(0),z_h}_{g}$
reproduce the standard $\overline{\rm MS}$ DIS coefficient functions $f_{q,2}$
and $f_{g,2}$ of $\mathcal{F}_2$, as required by the momentum sum rule~\cite{Altarelli:1979ub}.The unweighted OPEC is
\begin{align}
\Sigma_{z_h}^{(0),T-2}
={}&\sigma_0\sum_{q,\bar q}Q_q^2\int_{x_B}^1\frac{dx}{x}\Bigg\{
f_{q/A}(x,\mu^2)\Big[\delta(1-\hat x)+\frac{\alpha_s}{2\pi}\Big(
\ln\frac{Q^2}{\mu^2}P_{qq}(\hat x)+H^{(0),z_h}_{q}(\hat x)\Big)\Big]
\nonumber\\
&\qquad+\frac{\alpha_s}{2\pi}f_{g/A}(x,\mu^2)\Big(
\ln\frac{Q^2}{\mu^2}P_{qg}(\hat x)+H^{(0),z_h}_{g}(\hat x)\Big)\Bigg\}.
\label{eq:opec_unweighted}
\end{align}

\paragraph{Weighted OPEC.}
Since $w(\hat x,1)=w(1,\hat z)=0$, the Born term, all
$\delta(1-\hat x)$ and $\delta(1-\hat z)$ terms, the plus-prescription
subtractions and all explicit collinear logarithms vanish in both
channels. Only the regular parts of $H^{T-2}_{qq}$, $H^{T-2}_{qg}$ and
$H^{T-2}_{gq}$ contribute:
\begin{align}
H^{(w),z_h}_{q}(\hat x)&=C_F\Bigg\{
-\frac{8\hat x^3(2\hat x^2-4\hat x+3)}{(2\hat x-1)^4}\,L
-\frac{2(8\hat x^4-24\hat x^3+34\hat x^2-13\hat x+2)}{(2\hat x-1)^3}
\nonumber\\
&\qquad+\kappa_y\Bigg[
\frac{8\hat x^2(1-\hat x)^2(1+\hat x)}{(2\hat x-1)^4}\,L
+\frac{4\hat x^2(1-\hat x)(5-4\hat x)}{(2\hat x-1)^3}\Bigg]\Bigg\},
\label{eq:hEw_q}
\\
H^{(w),z_h}_{g}(\hat x)&=T_R\Bigg\{
\frac{-4\hat x(1-\hat x)(8\hat x^4-16\hat x^3+12\hat x^2-6\hat x+3)}{(2\hat x-1)^4}\,L
-\frac{4(1-\hat x)(8\hat x^4-16\hat x^3+10\hat x^2-\hat x+1)}{(2\hat x-1)^3}
\nonumber\\
&\qquad+\kappa_y\Bigg[
\frac{16\hat x^2(1-\hat x)^3(\hat x+1)}{(2\hat x-1)^4}\,L
+\frac{8\hat x^2(1-\hat x)^2(5-4\hat x)}{(2\hat x-1)^3}\Bigg]\Bigg\}.
\label{eq:hEw_g}
\end{align}
As for the OPCC, the apparent poles at $\hat x=1/2$ cancel between the
logarithmic and rational terms. The weighted OPEC is
\begin{equation}
\Sigma_{z_h}^{(w),T-2}=\sigma_0\frac{\alpha_s}{2\pi}\sum_{q,\bar q}Q_q^2
\int_{x_B}^1\frac{dx}{x}\Big[f_{q/A}(x,\mu^2)\,H^{(w),z_h}_{q}(\hat x)
+f_{g/A}(x,\mu^2)\,H^{(w),z_h}_{g}(\hat x)\Big].
\label{eq:opec_weighted}
\end{equation}

\paragraph{Angular spread.}
\begin{equation}
\left\langle\sin^2\theta\right\rangle^{\Sigma_{z_h}}
=\frac{\Sigma_{z_h}^{(w),T-2}}{\Sigma_{z_h}^{(0),T-2}}.
\label{eq:angular_spread_opec}
\end{equation}
Unlike the OPCC, the OPEC receives a gluon-initiated contribution
through $f_{g/A}$, which makes it sensitive to the nuclear gluon density.

%====================================================================================================
\subsection{Angular-weighted OPEC and OPCC at twist 4}
\label{sec:angular_t4}

\paragraph{From $\ell_{hT}^2$ to $\sin^2\theta$ weighting.}
The twist-4 hard parts of Ref.~\cite{Kang:2014ela} are computed with the
weight $\ell_{hT}^2$, with an overall factor $z_h^2$ pulled out of the
master formula. We denote these $\ell_{hT}^2$-weighted hard parts by
$\mathcal H(\hat x,\hat z)$. For massless partons on the $2\to2$ shell, the
Breit-frame angle $\theta$ satisfies
\begin{equation}
\sin^2\theta=\mathcal{R}(\hat x,\hat z)\,\frac{\ell_{hT}^2}{z_h^2},\qquad
\mathcal{R}(\hat x,\hat z)\equiv\frac{4}{Q^2}\,
\frac{\hat x^2\hat z^2}{(1-\hat x-\hat z+2\hat x\hat z)^2}.
\label{eq:R}
\end{equation}
With $\ell_{hT}^2/z_h^2=Q^2(1-\hat x)(1-\hat z)/(\hat x\hat z)$, this
reproduces the twist-2 weight of eq.~\eqref{eq:angular_weight}. The
weight depends only on the momentum of the measured parton. The $k_\perp$
derivatives of the collinear expansion can therefore be traded for $x$
derivatives at fixed final-state momentum, and they never act on the
weight. The $\sin^2\theta$-weighted hard parts are then
\begin{equation}
\mathcal{R}(\hat x,\hat z)\,\mathcal H(\hat x,\hat z).
\label{eq:H_from_R}
\end{equation}
For the terms supported at $\hat x=1$ or $\hat z=1$, where $\ell_T\to0$,
the weight is differentiated rather than evaluated. Both weights are
quadratic in $\ell_T$ there, and since
\begin{equation}
\mathcal{R}(1,\hat z)=\mathcal{R}(\hat x,1)=\frac{4}{Q^2},
\end{equation}
eq.~\eqref{eq:H_from_R} still holds. All $\delta(1-\hat x)$ and
$\delta(1-\hat z)$ terms, including the virtual and divergent ones, are
thus multiplied by $4/Q^2$. The cancellation of the final-state
divergence through the sum rules, eq.~\eqref{eq-divT4}, is therefore
unaffected. Products with plus distributions reduce as
\begin{align}
\frac{\mathcal{R}}{(1-\hat x)_+(1-\hat z)_+}&=\frac{4}{Q^2}\,
\frac{1}{(1-\hat x)_+(1-\hat z)_+}
+\frac{\mathcal{R}-4/Q^2}{(1-\hat x)(1-\hat z)},\\
\frac{\mathcal{R}}{(1-\hat z)_+}&=\frac{4}{Q^2}\,\frac{1}{(1-\hat z)_+}
+\frac{\mathcal{R}-4/Q^2}{1-\hat z},
\end{align}
and likewise for $(1-\hat x)_+$. The second terms are regular, since
\begin{equation}
\mathcal{R}-\frac{4}{Q^2}=-\frac{4}{Q^2}\,
\frac{(1-\hat x)(1-\hat z)(1-\hat x-\hat z+3\hat x\hat z)}{(1-\hat x-\hat z+2\hat x\hat z)^2}
\end{equation}
vanishes linearly at both $\hat x=1$ and $\hat z=1$. In the following,
$\tilde H^{(w)}$ denotes the angular-weighted coefficients after the
$\hat z$ integration.
\paragraph{Range of validity and terms at $\hat x=1$.}
The twist-4 collinear expansion treats the rescattering kick $k_\perp$ as small
compared with the scale that sets the measured angle, namely the energy of the
measured parton, $\hat zQ/2$. For $\hat x=\hat z=1$ the measured parton carries the
full energy, and the expansion in $k_\perp^2/Q^2$ is well justified; this applies to
the LO term, the virtual correction, and the real-emission terms proportional to
$\delta(1-\hat x)\delta(1-\hat z)$. For $\hat x=1$ and $\hat z<1$, by contrast, the
measured parton can be soft, $\hat zQ\lesssim k_\perp$, in which case a single kick
changes its direction by an amount of order one and the expansion in
$k_\perp^2/(\hat z^2Q^2)$ breaks down. In the numerical results of the main text we
therefore keep, for both the OPEC and the OPCC, the LO, virtual and
$\delta(1-\hat x)\delta(1-\hat z)$ contributions, and omit the
$\mathcal O(\alpha_s)$ real-emission terms proportional to $\delta(1-\hat x)$ with
$\hat z<1$, with the plus distributions in $\hat z$ defined on $[0,1]$ as in
Ref.~\cite{Kang:2014ela}. The $\delta(1-\hat x)$ coefficients listed
below for OPEC and OPCC include both types of
contribution and are given for completeness; they are finite remainders, obtained
with a cut $\hat z\ge\hat z_{\min}$ after dropping the terms that depend on$\hat z_{\min}$.
%===================================================================
\subsection{Angular-weighted OPEC at twist 4}
\label{sec:angular_opec_t4}
We follow the twist-2 sections. The $\ell_{hT}^2$-weighted twist-4 finite
parts of Ref.~\cite{Kang:2014ela} are denoted
$\mathcal H^{T-4,ab}_{ij}(\hat x,\hat z)$. Here $i$ is the parton entering the
twist-4 correlator ($q$ for $T_{qg}$, $g$ for $T_{gg}$), $j$ is the
fragmenting parton, and $ab$ labels the scattering ($ss$, $hh$, $sh$, $hs$)
or the asymmetric cut ($L$, $R$). Their $\sin^2\theta$-weighted versions are
$\mathcal R\,\mathcal H^{T-4,ab}_{ij}$, as in eq.~\eqref{eq:H_from_R}, and for
the OPEC we define
\begin{equation}
\tilde H^{(w),z_h}_{ij,ab}(\hat x)\equiv\frac{Q^2}{4}\int_0^1 d\hat z\;\hat z\,
\mathcal R\,\mathcal H^{T-4,ab}_{ij}(\hat x,\hat z).
\label{eq:Htelda}
\end{equation}
When a channel contains several matrix-element structures, the
superscript $[n]$ labels them as in eqs.~\eqref{eq:opec_t4_Cqj} and
\eqref{eq:opec_t4_Cgq}. As before, $L\equiv\ln\frac{1-\hat x}{\hat x}$,
$\kappa_y=\frac{1+4(1-y)+(1-y)^2}{1+(1-y)^2}$, and
\begin{equation}
M\otimes G\equiv\int_{x_B}^1\frac{dx}{x}\,M(x)\,G(x_B/x).
\end{equation}
\paragraph{Weighted OPEC at twist 4.}
\begin{equation}
\Sigma^{(w),T-4}_{z_h}=\frac{16\pi^2\alpha_s\,\sigma_0}{N_c\,Q^2}\sum_{q,\bar q}Q_q^2
\Big\{T_{qg}(x_B,0,0)+\frac{\alpha_s}{2\pi}\Big[\mathcal C_{qq}+\mathcal C_{qg}+\mathcal C_{gq}+\mathcal C_{V}+\mathcal C_{2}\Big]\Big\},
\label{eq:opec_t4_assembly}
\end{equation}
where $\sigma_0=2\pi\alpha_{\rm em}^2[1+(1-y)^2]/(yQ^2)$ is the Born
normalization of the twist-2 sections, the matrix elements in the
$\mathcal C$'s carry the flavor $q$, and $\alpha_s$ and the factorization
scale are evaluated at $\mu=Q$. The first term in the braces is the
leading-order twist-4 contribution. For quark ($j=q$) and gluon ($j=g$) fragmentation,
\begin{align}
\mathcal C_{qj}={}&x^2\frac{d^2T_{qg}(x,0,0)}{dx^2}\otimes\tilde H^{(w),z_h[2]}_{qj,ss}
-x\frac{dT_{qg}(x,0,0)}{dx}\otimes\tilde H^{(w),z_h[1]}_{qj,ss}
+T_{qg}(x,0,0)\otimes\tilde H^{(w),z_h[0]}_{qj,ss}
\nonumber\\
&+T_{qg}(x_B,x-x_B,0)\otimes\tilde H^{(w),z_h}_{qj,hh}
\nonumber\\
&+x\frac{dT_{qg}(x,0,x_B-x)}{dx}\otimes\tilde H^{(w),z_h[1]}_{qj,sh}
+T_{qg}(x,0,x_B-x)\otimes\tilde H^{(w),z_h[0]}_{qj,sh}
\nonumber\\
&-x\frac{\partial T_{qg}(x,x_2,x_B-x)}{\partial x_2}\Big|_{x_2=0}\otimes\tilde H^{(w),z_h[2]}_{qj,sh}
\nonumber\\
&+x\frac{dT_{qg}(x_B,x-x_B,x-x_B)}{dx}\otimes\tilde H^{(w),z_h[1]}_{qj,sh}
+T_{qg}(x_B,x-x_B,x-x_B)\otimes\tilde H^{(w),z_h[0]}_{qj,sh}
\nonumber\\
&-x\frac{\partial T_{qg}(x_B,x_2,x-x_B)}{\partial x_2}\Big|_{x_2=x-x_B}\otimes\tilde H^{(w),z_h[2]}_{qj,sh}
\nonumber\\
&+T^L_{qg}(x,0,x_B-x)\otimes\tilde H^{(w),z_h[0]}_{qj,L}
-x\frac{\partial T^L_{qg}(x,x_2,x_B-x)}{\partial x_2}\Big|_{x_2=0}\otimes\tilde H^{(w),z_h[2]}_{qj,L}
\nonumber\\
&+T^R_{qg}(x_B,x-x_B,x-x_B)\otimes\tilde H^{(w),z_h[0]}_{qj,L}
\nonumber\\
&-x\frac{\partial T^R_{qg}(x_B,x_2,x-x_B)}{\partial x_2}\Big|_{x_2=x-x_B}\otimes\tilde H^{(w),z_h[2]}_{qj,L}.
\label{eq:opec_t4_Cqj}
\end{align}
The hard--soft and right-cut coefficients are equal to the soft--hard and
left-cut ones; only the matrix elements differ. The asymmetric-cut
coefficients $\tilde H_{qj,L}$ include the overall sign with which these
cuts enter the cross section, so that all terms in eq.~\eqref{eq:opec_t4_Cqj}
are added. The gluon-initiated term is
\begin{equation}
\mathcal C_{gq}=x^2\frac{d^2T_{gg}(x,0,0)}{dx^2}\otimes\tilde H^{(w),z_h[2]}_{gq}
-x\frac{dT_{gg}(x,0,0)}{dx}\otimes\tilde H^{(w),z_h[1]}_{gq}
-T_{gg}(x,0,0)\otimes\tilde H^{(w),z_h[0]}_{gq},
\label{eq:opec_t4_Cgq}
\end{equation}
the finite part of the virtual correction gives
\begin{equation}
\mathcal C_{V}=-8C_F\,T_{qg}(x_B,0,0),
\end{equation}
and the gluon-fragmentation channels contain, in addition,
\begin{align}
\mathcal C_{2}={}&2C_A\int_{x_B}^1\frac{dx}{x}\,\frac{1}{(1-x_B/x)^2}
\Big[T_{qg}(x,0,0)+T_{qg}(x_B,x-x_B,0)
-T_{qg}(x,0,x_B-x)-T_{qg}(x_B,x-x_B,x-x_B)\Big].
\label{eq:opec_t4_C2}
\end{align}
The bracket in eq.~\eqref{eq:opec_t4_C2} and its first derivative vanish
at $x=x_B$, so the integral converges.

\paragraph{Quark fragmentation: central cuts.}
\begin{align}
\tilde H^{(w),z_h[2]}_{qq,ss}={}&C_{F}\Big\{\frac{(1-\hat x) \hat x^{2} (4 \hat x^{2} - 2 \hat x - 1)}{(2 \hat x - 1)^{3}}L
- \frac{(1-\hat x) (4 \hat x^{4} + \hat x^{2} - 2 \hat x + 1)}{(2 \hat x - 1)^{2}}+\kappa_y\Big[\frac{4 (1-\hat x)^{3} \hat x^{3}}{(2 \hat x - 1)^{3}}L+\frac{2 (1-\hat x)^{2} \hat x^{2}}{(2 \hat x - 1)^{2}}\Big]\Big\}.
\label{eq:opec4_qss2}
\end{align}

\begin{align}
\tilde H^{(w),z_h[1]}_{qq,ss}&=C_{F}\Big\{- \frac{\hat x^{2} (8 \hat x^{3} - 16 \hat x^{2} + 6 \hat x + 1)}{(2 \hat x - 1)^{3}}L+\frac{2 (8 \hat x^{5} - 16 \hat x^{4} + 2 \hat x^{3} + 3 \hat x^{2} + \hat x - 1)}{(2 \hat x - 1)^{2}}
\nonumber\\&\quad+\kappa_y\Big[\frac{4 (1-\hat x)^{2} \hat x^{3} (4 \hat x - 3)}{(2 \hat x - 1)^{3}}L+\frac{2 (1-\hat x) \hat x^{2} (4 \hat x - 3)}{(2 \hat x - 1)^{2}}\Big]\Big\}.
\label{eq:opec4_qss1}
\end{align}

\begin{align}
\tilde H^{(w),z_h[0]}_{qq,ss}&=C_{F}\Big\{\frac{(1-\hat x) (8 \hat x^{4} - 8 \hat x^{3} + 2 \hat x^{2} - 4 \hat x + 1)}{(2 \hat x - 1)^{3}}L-\frac{\ln\hat x}{1-\hat x}+\frac{2 (12 \hat x^{5} - 28 \hat x^{4} + 15 \hat x^{3} + 8 \hat x^{2} - 10 \hat x + 2)}{(2 \hat x - 1)^{2}}
\nonumber\\&\quad+\kappa_y\Big[- \frac{8 (1-\hat x)^{2} \hat x^{3} (3 \hat x - 2)}{(2 \hat x - 1)^{3}}L- \frac{4 (1-\hat x) \hat x^{2} (3 \hat x - 2)}{(2 \hat x - 1)^{2}}\Big]
\nonumber\\&\quad-6\Big[\frac{1}{1-\hat x}\Big]_++\Big[\frac{\ln(1-\hat x)}{1-\hat x}\Big]_++\Big(7 - \frac{2 \pi^{2}}{3}\Big)\delta(1-\hat x)\Big\}.
\label{eq:opec4_qss0}
\end{align}

\begin{align}
\tilde H^{(w),z_h}_{qq,hh}&=C_{F}\Big\{- \frac{30 \hat x^{4} - 51 \hat x^{3} + 33 \hat x^{2} - 9 \hat x + 1}{(2 \hat x - 1)^{5}}L+\frac{\ln\hat x}{1-\hat x}- \frac{120 \hat x^{3} - 158 \hat x^{2} + 77 \hat x - 11}{6 (2 \hat x - 1)^{4}}
\nonumber\\&\quad- \frac{11}{6}\Big[\frac{1}{1-\hat x}\Big]_+-\Big[\frac{\ln(1-\hat x)}{1-\hat x}\Big]_++\Big(- \frac{\pi^{2}}{3} - \frac{1}{9}\Big)\delta(1-\hat x)\Big\}
\nonumber\\&\quad+C_{A}\Big\{\frac{17 \hat x^{3} - 27 \hat x^{2} + 14 \hat x - 2}{(2 \hat x - 1)^{4}}L-2\frac{\ln\hat x}{1-\hat x}+\frac{(1-\hat x) (4 \hat x + 1)}{2 (2 \hat x - 1)^{3}}
\nonumber\\&\quad- \frac{3}{2}\Big[\frac{1}{1-\hat x}\Big]_++2\Big[\frac{\ln(1-\hat x)}{1-\hat x}\Big]_++\Big(\frac{7}{2} - \frac{\pi^{2}}{3}\Big)\delta(1-\hat x)\Big\}.
\label{eq:opec4_qhh}
\end{align}

\begin{align}
\tilde H^{(w),z_h[1]}_{qq,sh}&=C_{F}\Big\{- \frac{\hat x^{2} (3 \hat x^{3} - 2 \hat x^{2} - \hat x + 1)}{(2 \hat x - 1)^{4}}L+\frac{\hat x (4 \hat x^{3} - 9 \hat x^{2} + 6 \hat x - 2)}{2 (2 \hat x - 1)^{3}}\Big\}
\nonumber\\&\quad+C_{A}\Big\{\frac{\hat x^{2} (11 \hat x^{3} - 12 \hat x^{2} + 2 \hat x + 1)}{2 (2 \hat x - 1)^{4}}L- \frac{28 \hat x^{4} - 13 \hat x^{3} - 22 \hat x^{2} + 18 \hat x - 4}{4 (2 \hat x - 1)^{3}}\Big\}.
\label{eq:opec4_qsh1}
\end{align}

\begin{align}
\tilde H^{(w),z_h[2]}_{qq,sh}&=C_{F}\Big\{- \frac{\hat x^{2} (2 \hat x^{4} - 3 \hat x^{3} + 4 \hat x^{2} - 3 \hat x + 1)}{(2 \hat x - 1)^{5}}L- \frac{\hat x^{2} (8 \hat x^{3} + 22 \hat x^{2} - 31 \hat x + 12)}{6 (2 \hat x - 1)^{4}}\Big\}
\nonumber\\&\quad+C_{A}\Big\{\frac{(1-\hat x) \hat x^{2} (2 \hat x^{3} - 2 \hat x + 1)}{(2 \hat x - 1)^{5}}L+\frac{\hat x (16 \hat x^{4} - 22 \hat x^{3} + 16 \hat x^{2} - 9 \hat x + 3)}{6 (2 \hat x - 1)^{4}}\Big\}.
\label{eq:opec4_qsh2}
\end{align}

\begin{align}
\tilde H^{(w),z_h[0]}_{qq,sh}&=C_{F}\Big\{\frac{3 \hat x^{5} - 9 \hat x^{4} + 18 \hat x^{3} - 16 \hat x^{2} + 7 \hat x - 1}{(2 \hat x - 1)^{4}}L-\frac{\ln\hat x}{1-\hat x}- \frac{(\hat x + 1) (4 \hat x^{3} - 9 \hat x^{2} + 4 \hat x - 1)}{2 (2 \hat x - 1)^{3}}
\nonumber\\&\quad+\frac{1}{2}\Big[\frac{1}{1-\hat x}\Big]_++\Big[\frac{\ln(1-\hat x)}{1-\hat x}\Big]_++\Big(\frac{\pi^{2}}{3} + \frac{31}{6}\Big)\delta(1-\hat x)\Big\}
\nonumber\\&\quad+C_{A}\Big\{- \frac{11 \hat x^{5} - 27 \hat x^{4} + 39 \hat x^{3} - 33 \hat x^{2} + 14 \hat x - 2}{2 (2 \hat x - 1)^{4}}L+\frac{\ln\hat x}{1-\hat x}+\frac{28 \hat x^{4} - 81 \hat x^{3} + 83 \hat x^{2} - 37 \hat x + 5}{4 (2 \hat x - 1)^{3}}
\nonumber\\&\quad+\frac{11}{4}\Big[\frac{1}{1-\hat x}\Big]_+-\Big[\frac{\ln(1-\hat x)}{1-\hat x}\Big]_++\Big(- \frac{67}{12} + \frac{\pi^{2}}{6}\Big)\delta(1-\hat x)\Big\}.
\label{eq:opec4_qsh0}
\end{align}

\paragraph{Quark fragmentation: asymmetric cuts.}
These coefficients include the overall minus sign with which the
asymmetric cuts enter the cross section in Ref.~\cite{Kang:2014ela}.
\begin{align}
\tilde H^{(w),z_h[2]}_{qq,L}&=C_{F}\Big\{\frac{\hat x^{2} (2 \hat x^{4} - 3 \hat x^{3} + 4 \hat x^{2} - 3 \hat x + 1)}{(2 \hat x - 1)^{5}}L+\frac{\hat x^{2} (8 \hat x^{3} + 22 \hat x^{2} - 31 \hat x + 12)}{6 (2 \hat x - 1)^{4}}\Big\}
\nonumber\\&\quad-C_{A}\Big\{\frac{(1-\hat x) \hat x^{2} (2 \hat x^{3} - 2 \hat x + 1)}{(2 \hat x - 1)^{5}}L+\frac{\hat x (16 \hat x^{4} - 22 \hat x^{3} + 16 \hat x^{2} - 9 \hat x + 3)}{6 (2 \hat x - 1)^{4}}\Big\}.
\label{eq:opec4_qL2}
\end{align}

\begin{align}
\tilde H^{(w),z_h[0]}_{qq,L}&=C_{F}\Big\{-\frac{5}{3}\,\delta(1-\hat x)\Big\}+C_{A}\Big\{\frac{4}{3}\,\delta(1-\hat x)\Big\}.
\label{eq:opec4_qL0}
\end{align}

\paragraph{Gluon fragmentation: central cuts.}
\begin{align}
\tilde H^{(w),z_h[2]}_{qg,ss}&=C_{A}\Big\{\frac{2 (1-\hat x) \hat x^{2} (2 \hat x^{2} - 2 \hat x + 1)}{(2 \hat x - 1)^{3}}L+\frac{\hat x (4 \hat x^{4} - 8 \hat x^{3} + 11 \hat x^{2} - 7 \hat x + 2)}{(2 \hat x - 1)^{2}}
\nonumber\\&\quad+\kappa_y\Big[- \frac{2 (1-\hat x)^{2} \hat x^{3}}{(2 \hat x - 1)^{3}}L- \frac{4 (1-\hat x)^{2} \hat x^{3}}{(2 \hat x - 1)^{2}}\Big]\Big\}.
\label{eq:opec4_gss2}
\end{align}

\begin{align}
\tilde H^{(w),z_h[1]}_{qg,ss}&=C_{A}\Big\{\frac{2 \hat x^{2} (4 \hat x^{3} - 8 \hat x^{2} + 4 \hat x - 1)}{(2 \hat x - 1)^{3}}L- \frac{16 \hat x^{5} - 28 \hat x^{4} + 20 \hat x^{3} - 11 \hat x^{2} + 8 \hat x - 2}{(2 \hat x - 1)^{2}}
\nonumber\\&\quad+\kappa_y\Big[- \frac{2 (1-\hat x) \hat x^{3} (4 \hat x - 3)}{(2 \hat x - 1)^{3}}L- \frac{4 (1-\hat x) \hat x^{3} (4 \hat x - 3)}{(2 \hat x - 1)^{2}}\Big]-2\Big[\frac{1}{1-\hat x}\Big]_+\Big\}.
\label{eq:opec4_gss1}
\end{align}

\begin{align}
\tilde H^{(w),z_h[0]}_{qg,ss}&=C_{A}\Big\{- \frac{2 (4 \hat x^{5} - 8 \hat x^{4} + 5 \hat x^{3} - 8 \hat x^{2} + 5 \hat x - 1)}{(2 \hat x - 1)^{3}}L-2\frac{\ln\hat x}{1-\hat x}+\frac{24 \hat x^{5} - 40 \hat x^{4} + 26 \hat x^{3} - 8 \hat x^{2} + 9 \hat x - 3}{(2 \hat x - 1)^{2}}
\nonumber\\&\quad+\kappa_y\Big[\frac{4 (1-\hat x) \hat x^{3} (3 \hat x - 2)}{(2 \hat x - 1)^{3}}L+\frac{8 (1-\hat x) \hat x^{3} (3 \hat x - 2)}{(2 \hat x - 1)^{2}}\Big]
\nonumber\\&\quad+\Big[\frac{1}{1-\hat x}\Big]_++2\Big[\frac{\ln(1-\hat x)}{1-\hat x}\Big]_++\Big(-7 + \frac{2 \pi^{2}}{3}\Big)\delta(1-\hat x)\Big\}.
\label{eq:opec4_gss0}
\end{align}

\begin{align}
\tilde H^{(w),z_h}_{qg,hh}&=C_{F}\Big\{\frac{2 \hat x^{2} (4 \hat x^{2} - 3 \hat x + 1)}{(2 \hat x - 1)^{5}}L+\frac{72 \hat x^{3} - 76 \hat x^{2} + 31 \hat x - 4}{3 (2 \hat x - 1)^{4}}
+\frac{4}{3}\Big[\frac{1}{1-\hat x}\Big]_+-\frac{43}{18}\,\delta(1-\hat x)\Big\}
\nonumber\\&\quad+C_{A}\Big\{\frac{(3 \hat x - 1) (19 \hat x^{2} - 16 \hat x + 4)}{(2 \hat x - 1)^{4}}L-4\frac{\ln\hat x}{1-\hat x}+\frac{76 \hat x^{2} - 61 \hat x + 13}{2 (2 \hat x - 1)^{3}}
\nonumber\\&\quad+\frac{9}{2}\Big[\frac{1}{1-\hat x}\Big]_++4\Big[\frac{\ln(1-\hat x)}{1-\hat x}\Big]_++\Big(- \frac{13}{2} + \frac{4 \pi^{2}}{3}\Big)\delta(1-\hat x)\Big\}.
\label{eq:opec4_ghh}
\end{align}

\begin{align}
\tilde H^{(w),z_h[1]}_{qg,sh}&=C_{A}\Big\{\frac{\hat x^{2} (\hat x + 1) (7 \hat x^{2} - 5 \hat x + 1)}{2 (2 \hat x - 1)^{4}}L+\frac{4 \hat x^{4} - 31 \hat x^{3} + 68 \hat x^{2} - 42 \hat x + 8}{4 (2 \hat x - 1)^{3}}
+2\Big[\frac{1}{1-\hat x}\Big]_+\Big\}.
\label{eq:opec4_gsh1}
\end{align}

\begin{align}
\tilde H^{(w),z_h[2]}_{qg,sh}&=C_{A}\Big\{\frac{\hat x^{2} (10 \hat x^{4} + 13 \hat x^{3} - 20 \hat x^{2} + 8 \hat x - 1)}{2 (2 \hat x - 1)^{5}}L+\frac{56 \hat x^{5} - 98 \hat x^{4} + 377 \hat x^{3} - 408 \hat x^{2} + 168 \hat x - 24}{12 (2 \hat x - 1)^{4}}+2\Big[\frac{1}{1-\hat x}\Big]_+\Big\}.
\label{eq:opec4_gsh2}
\end{align}

\begin{align}
\tilde H^{(w),z_h[0]}_{qg,sh}&=C_{A}\Big\{- \frac{7 \hat x^{5} - 9 \hat x^{4} + 87 \hat x^{3} - 101 \hat x^{2} + 42 \hat x - 6}{2 (2 \hat x - 1)^{4}}L+3\frac{\ln\hat x}{1-\hat x}- \frac{4 \hat x^{4} - 11 \hat x^{3} + 111 \hat x^{2} - 89 \hat x + 19}{4 (2 \hat x - 1)^{3}}
\nonumber\\&\quad- \frac{11}{4}\Big[\frac{1}{1-\hat x}\Big]_+-3\Big[\frac{\ln(1-\hat x)}{1-\hat x}\Big]_++\Big(\frac{5}{12} - \pi^{2}\Big)\delta(1-\hat x)\Big\}.
\label{eq:opec4_gsh0}
\end{align}

\paragraph{Gluon fragmentation: asymmetric cuts.}
As for quark fragmentation, these coefficients include the overall sign with which the asymmetric cuts enter the cross section.
\begin{align}
\tilde H^{(w),z_h[2]}_{qg,L}&=C_{A}\Big\{\frac{\hat x^{2} (10 \hat x^{4} + 13 \hat x^{3} - 20 \hat x^{2} + 8 \hat x - 1)}{2 (2 \hat x - 1)^{5}}L+\frac{56 \hat x^{5} - 98 \hat x^{4} + 377 \hat x^{3} - 408 \hat x^{2} + 168 \hat x - 24}{12 (2 \hat x - 1)^{4}}+2\Big[\frac{1}{1-\hat x}\Big]_+\Big\}.
\label{eq:opec4_gL2}
\end{align}

\begin{align}
\tilde H^{(w),z_h[0]}_{qg,L}&=C_{A}\Big\{-\frac{23}{6}\,\delta(1-\hat x)\Big\}.
\label{eq:opec4_gL0}
\end{align}

\paragraph{Gluon--gluon double scattering.}
\begin{align}
\tilde H^{(w),z_h[2]}_{gq}&=T_{R}\Big\{\frac{(1-\hat x)^{2} \hat x^{2} (8 \hat x^{4} - 16 \hat x^{3} + 20 \hat x^{2} - 18 \hat x + 7)}{(2 \hat x - 1)^{4}}L- \frac{(1-\hat x)^{2} (40 \hat x^{5} - 104 \hat x^{4} + 114 \hat x^{3} - 61 \hat x^{2} + 15 \hat x - 2)}{(2 \hat x - 1)^{3}}
\nonumber\\&\quad+\kappa_y\Big[- \frac{12 (1-\hat x)^{5} \hat x^{3}}{(2 \hat x - 1)^{4}}L+\frac{2 (1-\hat x)^{3} \hat x^{2} (4 \hat x^{2} - \hat x - 2)}{(2 \hat x - 1)^{3}}\Big]\Big\}.
\label{eq:opec4_gg2}
\end{align}

\begin{align}
\tilde H^{(w),z_h[1]}_{gq}&=T_{R}\Big\{- \frac{2 (1-\hat x) \hat x^{2}}{2 \hat x - 1}L- 48 \hat x^{4} + 110 \hat x^{3} - 86 \hat x^{2} + 26 \hat x - 3
\nonumber\\&\quad+\kappa_y\Big[- \frac{36 (1-\hat x)^{4} \hat x^{3}}{(2 \hat x - 1)^{3}}L+\frac{6 (1-\hat x)^{2} \hat x^{2} (4 \hat x^{2} - \hat x - 2)}{(2 \hat x - 1)^{2}}\Big]\Big\}.
\label{eq:opec4_gg1}
\end{align}

\begin{align}
\tilde H^{(w),z_h[0]}_{gq}&=T_{R}\Big\{\frac{(1-\hat x) \hat x^{2} (48 \hat x^{4} - 84 \hat x^{3} + 56 \hat x^{2} - 24 \hat x + 7)}{(2 \hat x - 1)^{3}}L- \frac{(1-\hat x) (240 \hat x^{5} - 588 \hat x^{4} + 538 \hat x^{3} - 225 \hat x^{2} + 41 \hat x - 3)}{(2 \hat x - 1)^{2}}
\nonumber\\&\quad+\kappa_y\Big[- \frac{12 (1-\hat x)^{3} \hat x^{3} (12 \hat x^{2} - 15 \hat x + 4)}{(2 \hat x - 1)^{4}}L+\frac{2 (1-\hat x) \hat x^{2} (4 \hat x^{2} - \hat x - 2) (12 \hat x^{2} - 15 \hat x + 4)}{(2 \hat x - 1)^{3}}\Big]\Big\}.
\label{eq:opec4_gg0}
\end{align}

The apparent poles at $\hat x=1/2$ cancel between the logarithmic and
rational terms in every coefficient.

%===================================================================
\subsection{Angular-weighted OPCC at twist 4: the $qq$ channel}
\label{sec:angular_opcc_t4}

\paragraph{Channels.}
For $\mathcal W_h=Q_h$, the sum rule of Eq.~\eqref{eq-sumrule} gives
$M_q=Q_q$, $M_{\bar q}=-Q_q$ and $M_g=0$. The gluon-fragmentation channels
therefore drop out. In the gluon-initiated term with $T_{gg}$, the quark and
the antiquark of each flavor enter with the same hard coefficient and
opposite $M_q$, and their contributions cancel. The twist-4 OPCC is thus
given entirely by the $qq$ channel: a quark from $T_{qg}$ fragmenting into
the measured hadrons, through the central cuts ($ss$, $hh$, $sh$, $hs$), the
asymmetric cuts ($L$, $R$), and the virtual correction.

\paragraph{Weighted OPCC at twist 4.}
Following the same notation, the weight carries no factor of $\hat z$ after the
change of variables, and we define
\begin{equation}
\tilde H^{(w),Q_h}_{qq,ab}(\hat x)\equiv\frac{Q^2}{4}\int_0^1 d\hat z\;\mathcal R\,
\mathcal H^{T-4,ab}_{qq}(\hat x,\hat z).
\label{eq:Htelda_opcc}
\end{equation}
The terms at $\hat x=1$ are treated as described in Sec.~\ref{sec:angular_t4}. Then
\begin{equation}
\Sigma^{(w),T-4}_{Q_h}=\frac{16\pi^2\alpha_s\,\sigma_0}{N_c\,Q^2}\sum_{q,\bar q}Q_q^2M_q
\Big\{T_{qg}(x_B,0,0)+\frac{\alpha_s}{2\pi}\Big[\mathcal C^{Q_h}_{qq}+\mathcal C_{V}+\mathcal C^{Q_h}_{2}\Big]\Big\},
\label{eq:opcc_t4_assembly}
\end{equation}
where $\mathcal C^{Q_h}_{qq}$ is given by Eq.~\eqref{eq:opec_t4_Cqj} with
$j=q$ and $\tilde H^{(w),z_h}\to\tilde H^{(w),Q_h}$, and
$\mathcal C_V=-8C_F\,T_{qg}(x_B,0,0)$ as for the OPEC. As there, the
hard--soft and right-cut coefficients equal the soft--hard and left-cut
ones, and the asymmetric-cut coefficients include the overall sign with
which these cuts enter the cross section. The terms proportional to
$[1/(1-\hat x)^2]_{++}$ in the $ss$, $hh$, $sh$ and $hs$ channels, with
coefficients $C_F$, $C_F$, $-C_F$ and $-C_F$, are collected into
\begin{align}
\mathcal C^{Q_h}_{2}={}&C_F\int_{x_B}^1\frac{dx}{x}\,\frac{1}{(1-x_B/x)^2}
\Big[T_{qg}(x,0,0)+T_{qg}(x_B,x-x_B,0)-T_{qg}(x,0,x_B-x)-T_{qg}(x_B,x-x_B,x-x_B)\Big],
\label{eq:opcc_t4_C2}
\end{align}
which converges for the same reason as Eq.~\eqref{eq:opec_t4_C2}: the
bracket and its first derivative vanish at $x=x_B$ for any $T_{qg}$.

\paragraph{Central cuts.}
\begin{align}
\tilde H^{(w),Q_h[2]}_{qq,ss}&=C_{F}\Big\{\frac{(1-\hat x) \hat x^{2}}{(2 \hat x - 1)^{2}}L+\frac{4 \hat x^{4} - 4 \hat x^{3} + 2 \hat x^{2} - 2 \hat x + 1}{2 \hat x - 1}+\kappa_y\Big[- \frac{2 (1-\hat x)^{2} \hat x^{3}}{(2 \hat x - 1)^{2}}L- \frac{2 (1-\hat x)^{2} \hat x^{2}}{2 \hat x - 1}\Big]\Big\}.
\label{eq:opcc4_ss2}
\end{align}

\begin{align}
\tilde H^{(w),Q_h[1]}_{qq,ss}&=C_{F}\Big\{\frac{\hat x^{2}}{(2 \hat x - 1)^{2}}L+\frac{16 \hat x^{4} - 22 \hat x^{3} - 3 \hat x^{2} + 3}{2 \hat x - 1}
\nonumber\\&\quad+\kappa_y\Big[- \frac{2 (1-\hat x) \hat x^{3} (4 \hat x - 3)}{(2 \hat x - 1)^{2}}L- \frac{2 (1-\hat x) \hat x^{2} (4 \hat x - 3)}{2 \hat x - 1}\Big]+\Big[\frac{1}{1-\hat x}\Big]_+\Big\}.
\label{eq:opcc4_ss1}
\end{align}

\begin{align}
\tilde H^{(w),Q_h[0]}_{qq,ss}&=C_{F}\Big\{\frac{3 \hat x - 1}{(2 \hat x - 1)^{2}}L-\frac{\ln\hat x}{1-\hat x}+\frac{24 \hat x^{4} - 36 \hat x^{3} + 6 \hat x^{2} + 11 \hat x - 3}{2 \hat x - 1}
\nonumber\\&\quad+\kappa_y\Big[\frac{4 (1-\hat x) \hat x^{3} (3 \hat x - 2)}{(2 \hat x - 1)^{2}}L+\frac{4 (1-\hat x) \hat x^{2} (3 \hat x - 2)}{2 \hat x - 1}\Big]-6\Big[\frac{1}{1-\hat x}\Big]_++\Big[\frac{\ln(1-\hat x)}{1-\hat x}\Big]_++\Big(4 - \frac{2 \pi^{2}}{3}\Big)\delta(1-\hat x)\Big\}.
\label{eq:opcc4_ss0}
\end{align}

\begin{align}
\tilde H^{(w),Q_h}_{qq,hh}&=C_{F}\Big\{\frac{13 \hat x^{3} - 15 \hat x^{2} + 7 \hat x - 1}{(2 \hat x - 1)^{4}}L-\frac{\ln\hat x}{1-\hat x}+\frac{20 \hat x^{2} - 13 \hat x + 3}{2 (2 \hat x - 1)^{3}}
\nonumber\\&\quad+\frac{1}{2}\Big[\frac{1}{1-\hat x}\Big]_++\Big[\frac{\ln(1-\hat x)}{1-\hat x}\Big]_++\Big(- \frac{1}{2} + \frac{\pi^{2}}{3}\Big)\delta(1-\hat x)\Big\}
\nonumber\\&\quad+C_{A}\Big\{- \frac{(1-\hat x) (4 \hat x - 1)}{(2 \hat x - 1)^{3}}L-\frac{\ln\hat x}{1-\hat x}- \frac{2 \hat x}{(2 \hat x - 1)^{2}}
\nonumber\\&\quad-2\Big[\frac{1}{1-\hat x}\Big]_++\Big[\frac{\ln(1-\hat x)}{1-\hat x}\Big]_++\Big(1 - \frac{2 \pi^{2}}{3}\Big)\delta(1-\hat x)\Big\}.
\label{eq:opcc4_hh}
\end{align}

\begin{align}
\tilde H^{(w),Q_h[1]}_{qq,sh}&=C_{F}\Big\{\frac{2 (1-\hat x) \hat x^{3}}{(2 \hat x - 1)^{3}}L+\frac{(1-\hat x) (3 \hat x - 1)}{(2 \hat x - 1)^{2}}+\Big[\frac{1}{1-\hat x}\Big]_+\Big\}
\nonumber\\&\quad+C_{A}\Big\{\frac{\hat x^{3} (4 \hat x - 3)}{2 (2 \hat x - 1)^{3}}L- \frac{3 \hat x^{3} + \hat x^{2} - 3 \hat x + 1}{(2 \hat x - 1)^{2}}\Big\}.
\label{eq:opcc4_sh1}
\end{align}

\begin{align}
\tilde H^{(w),Q_h[2]}_{qq,sh}&=C_{F}\Big\{- \frac{\hat x^{2} (\hat x^{3} - 4 \hat x^{2} + 3 \hat x - 1)}{(2 \hat x - 1)^{4}}L- \frac{4 \hat x^{4} + 3 \hat x^{3} - 16 \hat x^{2} + 10 \hat x - 2}{2 (2 \hat x - 1)^{3}}+\Big[\frac{1}{1-\hat x}\Big]_+\Big\}
\nonumber\\&\quad+C_{A}\Big\{- \frac{\hat x^{2} (3 \hat x^{3} - 2 \hat x^{2} - \hat x + 1)}{2 (2 \hat x - 1)^{4}}L+\frac{\hat x (4 \hat x^{3} - 9 \hat x^{2} + 6 \hat x - 2)}{4 (2 \hat x - 1)^{3}}\Big\}.
\label{eq:opcc4_sh2}
\end{align}

\begin{align}
\tilde H^{(w),Q_h[0]}_{qq,sh}&=C_{F}\Big\{\frac{2 \hat x^{3} (\hat x - 2)}{(2 \hat x - 1)^{3}}L- \frac{\hat x^{2} + 3 \hat x - 1}{(2 \hat x - 1)^{2}}+4\,\delta(1-\hat x)\Big\}
\nonumber\\&\quad+C_{A}\Big\{\frac{(1-\hat x) (4 \hat x^{3} - 3 \hat x^{2} + 4 \hat x - 1)}{2 (2 \hat x - 1)^{3}}L+\frac{1}{2}\frac{\ln\hat x}{1-\hat x}+\frac{6 \hat x^{3} - 16 \hat x^{2} + 14 \hat x - 3}{2 (2 \hat x - 1)^{2}}
\nonumber\\&\quad+3\Big[\frac{1}{1-\hat x}\Big]_+- \frac{1}{2}\Big[\frac{\ln(1-\hat x)}{1-\hat x}\Big]_++\Big(-3 + \frac{\pi^{2}}{3}\Big)\delta(1-\hat x)\Big\}.
\label{eq:opcc4_sh0}
\end{align}

\paragraph{Asymmetric cuts.}
\begin{align}
\tilde H^{(w),Q_h[2]}_{qq,L}&=C_{F}\Big\{\frac{\hat x^{2} (\hat x^{3} - 4 \hat x^{2} + 3 \hat x - 1)}{(2 \hat x - 1)^{4}}L+\frac{4 \hat x^{4} + 3 \hat x^{3} - 16 \hat x^{2} + 10 \hat x - 2}{2 (2 \hat x - 1)^{3}}-\Big[\frac{1}{1-\hat x}\Big]_+\Big\}
\nonumber\\&\quad+C_{A}\Big\{\frac{\hat x^{2} (3 \hat x^{3} - 2 \hat x^{2} - \hat x + 1)}{2 (2 \hat x - 1)^{4}}L- \frac{\hat x (4 \hat x^{3} - 9 \hat x^{2} + 6 \hat x - 2)}{4 (2 \hat x - 1)^{3}}\Big\}.
\label{eq:opcc4_L2}
\end{align}

\begin{align}
\tilde H^{(w),Q_h[0]}_{qq,L}&=C_{F}\Big\{-\delta(1-\hat x)\Big\}+C_{A}\Big\{\frac{1}{2}\,\delta(1-\hat x)\Big\}.
\label{eq:opcc4_L0}
\end{align}

\end{document}